# Visibility of Domain Elements in the Elicitation Process: A Family of Empirical Studies

Alejandrina Aranda, Oscar Dieste, José Ignacio Panach and Natalia Juristo

***Abstract—Background*: Various factors determine analyst effectiveness during elicitation. While the literature suggests that elicitation technique and time are influential factors, other attributes could also play a role. *Aim*: Determine aspects that may have an influence on analysts' ability to identify certain elements of the problem domain. *Methodology*: We conducted 14 quasi-experiments, inquiring 134 subjects about two problem domains. For each problem domain, we calculated whether the experimental subjects identified the problem domain elements (concepts, processes, and requirements), i.e., the degree to which these domain elements were visible. *Results*: Domain element visibility does not appear to be related to either analyst experience or analyst-client interaction. Domain element visibility depends on how analysts provide the elicited information: when asked about the knowledge acquired during elicitation, domain element visibility dramatically increases compared to the information they provide using a written report. *Conclusions*: Further research is required to replicate our results. However, the finding that analysts have difficulty reporting the information they have acquired is useful for identifying alternatives for improving the documentation of elicitation results. We found evidence that other issues, like domain complexity, the relative importance of different elements within the domain, and the interview script, also seem influential.**

## I. INTRODUCTION

Problem domain comprehension is key to requirements engineering (RE). Being able to manage problem domain knowledge is critical to the success of a project [1]. One of the major problems in RE is incomplete and/or hidden requirements [2]. This problem is caused, at least in part, by an inadequate understanding of the domain.

Comprehension is a process that gradually occurs throughout the RE process [3] or even at later stages of the development process. Comprehension is especially important during elicitation [4]-[8] and, to a lesser extent, analysis [7]. This research focuses on elicitation.

Our earlier research [9] revealed that using a short open-ended interview, a wide-ranging group of analysts with different experience levels acquire only a limited number of problem domain elements. In other words, the completeness of the elicited information is far below 100% of the domain elements. A priori, this is not surprising: it makes sense that problem domain size and available time will place constraints on the number of identified problem domain elements. Besides, the results of the elicitation techniques depend on the context in which they are used [10]. The empirical research in this regard is limited [11]. However, it has been observed, for example, that some techniques, like protocol analysis [12] or rank ordering [13] are not necessarily efficient in certain domains or for identifying some types of information.

A more in-depth analysis of the information collected in [6] reveals that analyst effectiveness is more complex than completeness. We observe that some domain elements are identified more often than others. For example, in a telephony domain (see Section IV), the "message" concept is discovered much more frequently than the "contact" concept. Therefore, completeness only partially describes the analyst's effectiveness; two analysts achieving similar completeness (say 40% of the elements of the problem domain) are equally effective only if they identify the same domain elements. If they identify different domain elements, effectiveness differs: The overall completeness of the two analysts together, in the first case, would be 40%, whereas in the second case, it could rise to 80% if both analysts do not identify common domain elements.

**This research aims to determine which problem domain elements are identified by elicitors and explore the reasons for differential identification.** We analysed the 12 quasi-experiments described in [6] at a fine-grained level[1], plus two new quasi-experiments (Q16a and Q16b, see Section 4.1) specifically designed for this research. In total, 134 subjects participated in the research, providing 148 interviews. The dataset is available in the online Appendix. For the research reported here, we calculated the visibility of the problem domain elements, i.e., the probability of an element being identified, and we statistically determined whether they were visible (P(visibility) >= 0.5) or not (P(visibility) >= 0.5). Finally, we explored diverse factors, e.g., the analyst's experience, the respondent's talkativeness, problem domain

Alejandrina Aranda, Oscar Dieste and Natalia Juristo are with the Escuela Técnica Superior de Ingenieros Informáticos of the Universidad Politécnica de Madrid, Campus de Montegancedo, 28660 Boadilla del Monte, Madrid, SPAIN (e-mail: alearanda@gmail.com, {odieste, natalia}@fi.upm.es). José Ignacio Panach is with the Escuela Técnica Superior de Ingeniería of the Universidad Politécnica de Valencia, Avinguda de la Universitat s/n 46100 Burjassot, (email: j.ignacio.panach@uv.es).

[1] In [9], we used Completeness (the number of elicited domain elements over the total number of domain elements) as the response variable. In this research, we analyse each domain element individually, calculating their Visibility and providing a more nuanced view of the analyst's effectiveness.

structure, etc., to determine the potential causes of the differential visibilities. The contributions of this paper are as follows:

- We found that domain familiarity or unfamiliarity, analyst experience, or respondent talkativeness do not significantly influence problem domain elements' visibility.
- We observed that documenting the results of an interview session in writing is difficult. Analysts recall but fail to write up many problem domain elements.
- Visibility seems influenced by the relative importance of each element within the domain.

As the interview is the elicitation technique used in [9] and Q16a/b, our results are not directly generalisable to any other elicitation technique. Note, however, that the interview is the most relevant elicitation technique used in RE [14].

This article is structured as follows. Section II describes the related work that is mentioned later in the discussion. Sections III and IV describe the aims and methodology, respectively. Sections V and VI report the results. Section VII offers a detailed discussion of the results considering the existing literature. Finally, Sections VIII and IX describe the validity threats and conclusions. A web appendix is provided.

## II. Related work

Personal characteristics, particularly domain knowledge and analyst experience, are the most studied aspects of interview effectiveness. Aranda et al. [15][9] reviewed the findings in this respect, which can be summarised briefly as 1) **the domain influences requirements elicitation** insofar as it is easier to acquire information in familiar domains; 2) **it is clear that experience** (defined not as domain knowledge but as the number of years in the profession) **influences elicitation effectiveness in familiar domains,** whereas **its influence in unfamiliar domains is uncertain**.

This research aims to explore why the elicitors in [9] identify a limited number of problem domain elements. Our research does not include instrumental covariates, such as templates or types of questions, to mention two examples; the explanation must, therefore, lie in the interaction between the interviewer, respondent and problem domain. Consequently, we reviewed the literature in search of empirical studies that use a **simple user-analyst interaction**, such as an unstructured interview or any of its variants. We also reviewed any studies where the analyst consults a natural language problem description instead of eliciting information from a user proper, as they are applicable for our purposes (see Section VII). We have not included either theoretical works or acquisition tools or methods, as they pursue different objectives from ours. The aim of the studies should be to determine **analyst effectiveness and the reasons behind problem-domain information acquisition**. Effectiveness can be defined in different terms (quantity, accuracy, completeness, etc.) and measured in different ways (number, percentage, recall, etc.). Therefore, it is quite difficult to identify this type of studies, and we **assume that the following are only a subset of relevant studies.**

**Agarwal and Tanniru** [16] conducted a study with 20 graduate students (unexperienced engineers) and 10 experienced practitioners (experienced engineers). The experimental task was to make a decision on capital budgeting/resource allocation, for which purpose it was necessary to elicit rules and decision-making criteria. The aim of the experiment was to compare the effectiveness (not the completeness) of unstructured vs. the structured interviews in terms of number of elicited criteria and rules. This type of study would not, in principle, be of interest. In this case, however, the structured interview was created using a business decision-making domain model, called Duncan's model. Therefore, this research indirectly studies the impact of analyst access to a domain model on interview effectiveness. **The experimental results suggest that the structured interview outperforms the unstructured interview in terms of the number of both elicited criteria and elicited rules.** Another interesting result already reported in [9] suggests that analyst experience does not influence their effectiveness to the point that inexperienced analysts using the structured interview outperformed experienced analysts using the unstructured interview.

**Kiris** [17] conducted an experiment in an area related to requirements elicitation: knowledge acquisition for expert systems. Three techniques (interview, verbal protocol and concept mapping) were applied in two different domains (libraries and aircraft piloting). The experimental subjects were librarians and pilots, who interacted as experts, and postgraduate students, who acted as knowledge engineers. Kiris [17] measured several dimensions of the elicitation process, of which knowledge base accuracy and completeness are of interest for our purposes. The research revealed that **the domain had a significant effect in terms of completeness but not with respect to accuracy**. No other factor had significant effects. Additionally, Kiris [17] suggests that **the quality of the acquired knowledge depends more on the personal characteristics of the experts and knowledge engineers than on the elicitation techniques**. The term "quality" is listed in the hypotheses of [17] as a possible synonym for accuracy and completeness rather than as a response variable. We infer that Kiris [17] is referring to accuracy and completeness when discussing quality.

**Wilt** [18] conducted an experiment comparing the laddered grid and verbal protocol analysis elicitation techniques in the declarative, procedural and mixed domains. A pilot study was carried out beforehand to define these domains. The experimental subjects were employees (no further specification is given) of a US company. This work is interesting because, apart from the two elicitation techniques, Wilt defines different types (declarative, procedural and ambiguous) of prompts that are used during the elicitation session. A prompt is, essentially, a type of question with a specific aim [19]. This concept has often been applied in the requirements field. Context-dependent interviews [20], for example, are actually a special type of prompt. Question taxonomies have even been proposed [21]. The specific prompts used in Wilt's research are outlined in Table 4 of [18]. As far as we are concerned, the most interesting point of this research is the analysis of the domain x prompt interaction. To do this, a response variable with a somewhat peculiar definition, called *knowledge proportion,* is used. This response variable can, however, be construed as a type of recall (see pp. 67 of [18]). **The experiment showed that the procedural prompt acquired more procedural knowledge in the declarative and mixed domains than the declarative or ambiguous prompts.** Interestingly, there were fewer differences between prompts in the procedural domain.

TABLE I. SUMMARY OF RELATED WORK

| Study | Goal | Main factor(s) | Other factor(s)/relevant aspects | Response variable(s) | Outcomes |
|---|---|---|---|---|---|
| **Agarwal and Tanniru** [16] | Make a decision on capital budgeting/resource allocation. | Structured and unstructured interview | The structured interview was created using a business decision domain model. | Number of elicited criteria or rules | The structured interview (using Duncan's model) outperforms the unstructured interview in terms of number of both elicited criteria and rules. |
| **Kiris** [17] | Acquire knowledge for expert systems. | Interview, verbal protocol and concept mapping | Two different domains (libraries and aircraft piloting) were used. | Accuracy and completeness | The domain had a significant effect for completeness but not for accuracy. |
| **Wilt** [18] | Elicit requirements. | Laddered grid and verbal protocol analysis in three different domains: declarative, procedural and mixed | Declarative, procedural and ambiguous prompts were used. | Knowledge proportion (a type of recall) | The experiment shows that the procedural prompt elicited more procedural knowledge in the declarative and mixed domains than the declarative or ambiguous prompts. |
| **Browne and Rogich** [22] | Identify requirements of an online grocery shop. | Task characteristics interview, semantic interview and syntactic interview | The types of questions used in the interviews are substantially different. | Number of elicited requirements and scope of requirements (how many different categories the elicited requirements belong to) | In terms of number of elicited requirements, the task characteristics interview outperforms the semantic interview, which, in turn, fares better than the syntactic interview. No significant differences were observed with respect to the scope. |
| **Pitts and Browne** [23] | Identify requirements of an online grocery shop. | Interviews | Procedural prompts or interrogatory prompts were used. | Quantity and completeness of elicited requirements | The experimental results suggest that the procedural prompts increase the quantity and completeness of the elicited requirements. |
| **Burnay et al.** [24] | Elicit requirements. | Interviews | Topic set (a set of related entities of the domain of discourse were used. | Information spontaneously mentioned or not mentioned by users | Users tend to spontaneously mention information related to some topic sets but not to others. |
| **Dalpiaz et al.** [25] | Generate static conceptual models. | User cases or user stories | Several use cases were used: urban traffic simulator, International Football Association portal, and hospital management system. | Validity and completeness of the resulting models | The initial notation is not statistically significant. The complexity of the case study is more influential and is almost significant. |

**Browne and Rogich** [22] conducted an experiment with 45 "non-faculty workers" (sic) with experience in computers and databases. The experimental task was to identify the requirements of an online grocery shop. To do this, the subjects applied three different interviews: task characteristics interview, semantic interview and syntactic interview. The effectiveness of the interviews was measured in two different ways: number of elicited requirements and scope of requirements (how many different categories the elicited requirements belong to). Like Agarwal and Tanniru's research [16], Browne and Rogich's study [22], which, ostensibly, merely compares interviews rather than investigating the reasons why one interview might be better than another, would not appear to be of interest. Besides, the three interview types are context independent. However, the question types used are substantially different, and this could provide some interesting insights. For a description of the interviews, which, for reasons of space, cannot be described here, see Tables 3-5 in [22]. The results of the experiment show that the task characteristics interview elicits more requirements than the semantic interview, which, in turn, outperforms the syntactic interview. With regard to scope, no significant differences were observed. In our opinion, the significant results are related to the generality of the questions of each interview type. **The syntactic interview contains overly abstract questions, for example, *Who uses the system?* Or *What kinds of things do users do?* On the other hand, the questions of the other interview types are more sophisticated, for example, *What would your customers want the system to do?*** While we are aware that we are not defining what we mean by generality, we prefer to use the intuitive concept rather than open up a debate that this not central to this research.

**Pitts and Browne** [23] conducted an experiment with 54 practising analysts with at least two years of requirements experience. The experimental task was to identify the requirements of an online grocery shopping information system. To do this, the experimental subjects conducted two interviews. During the first session, they used an unstructured interview, whereas, in the second session, they were given the choice between two types of semi-structured interviews similar to context-independent interviews: one contained procedural prompts and the other contained interrogatory prompts. They use a similar concept of prompt to Wilt [18]. The procedural prompts are quite specific questions that aim to avoid the cognitive obstacles facing analysts during elicitation, whereas the interrogatory prompts are general questions. To get a clearer idea of the meaning of these concepts, see Table 3 in [23]. The experiment determined the quantity and completeness of the elicited requirements. Completeness was measured from two viewpoints: 1) breadth (number of different requirements categories that were utilized) and 2) depth (number of requirements acquired within each requirements category). **The experimental results suggest that procedural prompts increase the quantity and completeness of the requirements elicited by analysts.**

**Burnay et al.** [24] reported three empirical studies on the concept of 'topic set'. They define a topic as an entity type relevant to the domain of discourse, such as roles, activities, rules, etc. A topic set is a set of related topics. Despite the obvious differences, the concept of topic set recalls Agarwal and Tanniru's structured interview [16] or Pitts and Browne's [23] and Browne and Rogich's [22]: a checklist that points to certain problem domain elements. In the first study, the researchers interviewed several requirements engineers and/or business analysts working on different software projects. As a result, they got a list of topics, organized by topic sets. This information was used in the second empirical study, which is of special interest to us (we do not discuss the third study). This second study consisted of a survey. The participants were professionals that had participated as users in software projects. These users were asked questions, such as *During an interview with the business analyst, would you mention 'X', where X is to be replaced by a topic?* They were also asked whether *in their own experience, the topic sets are discussed with business analysts during interviews*. The results of this second study were inconclusive: **users tend to spontaneously mention information related to some topic sets but not to others.** Another result of this second study was that more experienced users tend to spontaneously mention information more often.

**Dalpiaz et al.** [25] conducted two empirical studies: a controlled experiment followed by a quasi-experiment using students as experimental subjects. The aim of the studies was to generate static conceptual models (probably class diagrams) based on two product types: use cases and user stories. They studied the validity and completeness of the resulting models. Validity was defined as the precision of the model generated by the subjects, whereas completeness was defined as recall. A model pattern was used to calculate precision and recall. The controlled experiment concluded that the conceptual models constructed from user stories were more valid and complete, albeit with some distinctions. The experiment used two case studies: data hub and planning poker. User stories only had a positive effect on the first object. As far as we are concerned, the follow-on quasi-experiment is of special interest. The experimental subjects performed the same experimental task, with two exceptions: (1) the experimental subjects interviewed a simulated user to create the use cases and user story, which were not provided by the researchers in this case, and (2) several case studies —urban traffic simulator, International Football Association portal, and hospital management system— were used. Although the validity and completeness variables were defined slightly differently, they are still comparable to precision and recall, respectively. The results suggest that the initial notation is not statistically significant, although the user stories are still slightly better than use cases. **Case study complexity, on the other hand, is more influential and almost significant.**

TABLE I summarizes the related work. While it is noticeably diverse, some trends can be observed:

- The use of some sort of domain model increases analyst effectiveness [16][26][24].
- The use of prompts designed to gather certain types of information increases analyst effectiveness [18][22][23].
- Analyst effectiveness depends on the type of domain [17][25]. Note that a similar result was already reported earlier for the dichotomy between familiar vs. unfamiliar domains [15].

## III. RESEARCH QUESTION

This research aims to gain insight into how analysts capture/acquire/elicit information. We conducted an exploratory study of the influence of problem domain elements on the elicitation process. To do this, we formulated the following research question:

**RQ**. Are all problem domain elements equally easy to identify at elicitation time?

Both common sense and professional experience suggest that not all elements are identified simultaneously, as they are not equally important to stakeholders. In a shopping portal, for example, products are a very important element, whereas means of payment are much less so. Additionally, analysts differ. The patterns for identifying the domain elements could vary depending on their personal characteristics: some elements could be identified immediately, whereas others may appear more or less randomly throughout the discourse.

In this research, we introduce the concept of *Visibility*. We define visibility as the extent to which a domain element appears in a **series of elicitation sessions**. From the mathematical viewpoint, an element's visibility is the ratio (or percentage) of sessions where an element is over the total number of elicitation sessions conducted. We have used unstructured interviews in this research. Therefore, visibility can also be defined as the number of elicitors identifying a problem domain element over the total number of elicitors. We have labelled specific ranges of visibility values:

- We regard an element as being *identified by chance* when the number of analysts who do and do not identify this element is more or less the same. In other words, the probability of identifying an element by chance is 50%.
- We regard an element as *easily observable* if the probability of being systematically identified is over 50%.
- We regard an element as *hard to observe* if the probability of it being systematically identified is under 50%.

This means that analysts tend to always (or often, over 50% of the time) identify visible elements, whereas they tend to always (or often, over 50%) overlook non-visible elements. Although this labelling is equivalent to a discretization of the visibility variable, there is no associated validity threat, as the statistical treatment is applied to the non-discretized variable. The labels are only used in the discussion of the results.

TABLE II. FAMILY OF EMPIRICAL STUDIES ON REQUIREMENTS ELICITATION

| Studies | | Empirical Study | Replication Type | No. Subjects | Analysts | Site | Elicitation Technique | Respondent | Language | Problem domain | #Exp. Units |
|---|---|---|---|---|---|---|---|---|---|---|---|
| 1 | Q07 | Baseline Quasi-experiment | Baseline experiment | 7 | MSc students | UPM | Interview | OD | English | Recycling | 7 |
| 2 | Q09 | Quasi-experiment | Internal replication Q07 | 8 | MSc students | UPM | Interview | AG | English | Recycling | 8 |
| 3 | Q11 | Quasi-experiment | Internal replication Q09 | 16 | MSc students | UPM | Interview | OD | English | Recycling | 16 |
| 4 | Q12 | Quasi-experiment | External replication Q11 | 21 | Researchers & practitioners | REFSQ | Interview | OD | English | Recycling | 21 |
| 5 | E12 | Experiment (Within Subjects) | Baseline experiment | 14 | MSc students | UPM | Interview | OD / JW | English / Spanish | Recycling Messaging | 28 |
| 6 | E13 | Experiment (Between Subjects) | Baseline experiment | 8<br>7 | MSc students | UPM | Interview | OD / JW | English / Spanish | Recycling Messaging | 15 |
| 7 | E14 | Experiment (Between Subjects) | Internal replication E13 | 9<br>7 | MSc students | UPM | Interview | OD | English | Recycling Messaging | 16 |
| 8 | E15 | Experiment (Between Subjects) | Internal replication E14 | 5<br>8 | MSc students | UPM | Interview | OD / JW | English / Spanish | Recycling Messaging | 13 |
| 9 | Q16a | Quasi-experiment | Internal replication E12 | 14 | MSc students | UPM | Listening to a domain description | - | English / Spanish | Recycling | 14 |
| 10 | Q16b | Quasi-experiment | Internal replication E12 | 10 | MSc students | UPM | Listening to a domain description | - | English | Messaging | 10 |
| **TOTAL** | | Studies with varying degrees of control | 10 empirical studies 14 quasi-experiments | 134 | Students and Practitioners | UPM + REFSQ | Interview / Recorded audio | 3 | 2 | 2 | 148<br>102 Recycling<br>46 Messaging |

## IV. RESEARCH METHOD

This section briefly describes the empirical studies and problem domains used in this research. Additionally, we describe the methodology and analysis process enacted to determine how visible the problem domain elements are.

### *A. Family of empirical studies*

We conducted ten empirical studies (including quasi-experiments and controlled experiments) to analyse analyst effectiveness in the requirements elicitation process. The **experimental subjects** participating in the studies ranged from Universidad Politécnica de Madrid postgraduate students taking a Requirements Engineering (RE) course to experienced professionals knowledgeable about computer science and related disciplines through researchers, faculty, and PhD students from several universities and research groups. TABLE II shows the conditions under which each study was conducted. We have 134 subjects, equivalent to 148 experimental units. The informed consents have been discussed in previous publications.

### *B. Problem domains*

We used two different problem domains: Messaging (Aware Problem - AP) and Battery Recycling (Ignorant problem - IP). AP refers to message sending/receiving using instant messaging applications, whereas IP is concerned with the battery recycling process. TABLE III shows both problem domains used in the empirical studies. Note that the IP was the most commonly used problem domain throughout the sequence of empirical studies, whereas AP was used in the experiments.

TABLE III. PROBLEM DOMAINS

| Domains | Empirical studies (Q – Quasi-experiment, E – Experiment) | | | | | | | | |
|---|---|---|---|---|---|---|---|---|---|
| | Q07 | Q09 | Q11 | Q12 | E12 | E13 | E14 | E15 | Q16 |
| AP | | | | | x | x | x | x | x |
| IP | x | x | x | x | x | x | x | x | x |

The problem domains used in this research were fully described in terms of the three types of elements: requirements, concepts, and processes, as shown in Appendix A. The problem domain elements were categorized as requirements, concepts, and processes after comparing how the different types of problem domain elements were categorized in the existing literature. Although the details of the studies reported in the literature differ, they do agree on the main elements. For example, according to [28], problems are usually analysed in terms of organizational goals, business processes, and tasks to be performed to achieve the goals. Likewise, [29] mentions that irrespective of the language, notation, or technique applied, requirements: 1) define an object, function, or state, 2) limit or control the actions associated with an object, function, or state, or 3) define relationships between objects, functions, or states. Apart from the above concepts, actions, or states, more modern formulations, like the Frisco Report [19], expressly account for elements implicit in the earlier literature, like actors or rules.

TABLE IV. TOTAL NUMBER OF DOMAIN ELEMENTS

| Problem domains | Elements defining the problem (#) | | | |
|---|---|---|---|---|
| | Requirements | Concepts | Processes | TOTAL |
| Messaging Domain | 28 | 10 | 16 | 54 |
| Recycling Domain | 15 | 24 | 12 | 51 |

Although originally considered, we later rejected goals, as they were hugely outnumbered by processes or concepts, and they did not provide key information for experimental purposes. We did not account for more sophisticated issues like actors and business rules, as the domains are rather simple (not many actors, few or no business rules). Considering that the experiments focus on the first contact between the analyst and client (early interviews), it is unrealistic to expect the analyst to be able to acquire fine details about the domain (inputs, outputs, attributes, relations, etc.). On this ground, we opted for a higher level of abstraction, and the above-mentioned details were not considered.

There are another two noteworthy issues: 1) both problems are based on real problems, which have been simplified to assure that they can be successfully tacked by MSc students in the limited time available for elicitation sessions; 2) an effort was made to define the different problems in such a way as to assure that the total number of problem elements was as similar as possible and to provide a full description of the problems using activity diagrams and concept models, as shown in Appendix A. TABLE IV shows the total number of elements defining and delimiting each of the problem domains.

### *C. Experimental task*

We defined visibility as the extent to which a domain element is observed in a set of elicitation sessions. This definition can be used in different settings:

- **Elicitation technique used**: The definition of visibility is independent of the technique type. It can be applied to interviews, focus groups, brainstorming, etc. In this research, we use the **unstructured interview**, partly because of the importance of interviews in RE [14] and partly because of our interests [15][9].
- **Elicitation session organization**: The sessions could be composed of a sequence of sessions carried out by the same analyst, that is, the first, second, third, etc., interviews of the same stakeholder. Another possibility is for more than one analyst to perform a single interview. Both approaches can be combined. In this research, we opted for the second option because 1) the conclusions of a single analyst would not appear to be generalisable, and 2) our previous experience (see [9], Appendix H) shows that interviews conducted one after another do not tend to identify new domain elements.

The **experimental task** was to enact a requirements elicitation process for one or more software systems corresponding to different problem domains. We used two problem domains (AP and IP) where the total number of subjects differed depending on the domain because not all the subjects elicited requirements for both domains. Therefore, of a total of 148 experimental units, 102 subjects correspond to the IP domain and 46 to the AP domain. We tested user domain familiarity by means of questionnaires (see [9], Table 7). We have used two domains where we had reason to think that **domain familiarity could affect the elicitation process**: common sense dictates that, in the familiar domain, the elements will be more visible than in the unfamiliar domain. However, our results (see Section V) do not confirm this hypothesis, although there is a possible explanation for this (see Section VI).

The **requirements elicitation process** was composed of two main phases: elicitation session and elicited information reporting. The elicitation session consisted of a 30-minute unstructured interview, during which the subjects acted as requirements analysts and a third party, sometimes but not always a research team member, played the client or respondent role. Three respondents (labelled with the initials OD, AG, and JW in TABLE II) participated throughout the research. To prevent the respondents from providing different information according to their understanding of the problem, a checklist was designed, including the elements defining the domains (AP, IP) under elicitation, and the respondents were trained to answer any questions posed by subjects as if they were real users/clients.

At the end of the elicitation session, the experimental subjects were given 90 minutes to write up all the information they had acquired during the interview in a report. The only exception was quasi-experiment Q12, where the elicitation and reporting times were 60 minutes. The reporting format was free in all cases (to prevent biases related to specific formats [30]). The researchers determined which problem domain elements the analysts had identified from the reports.

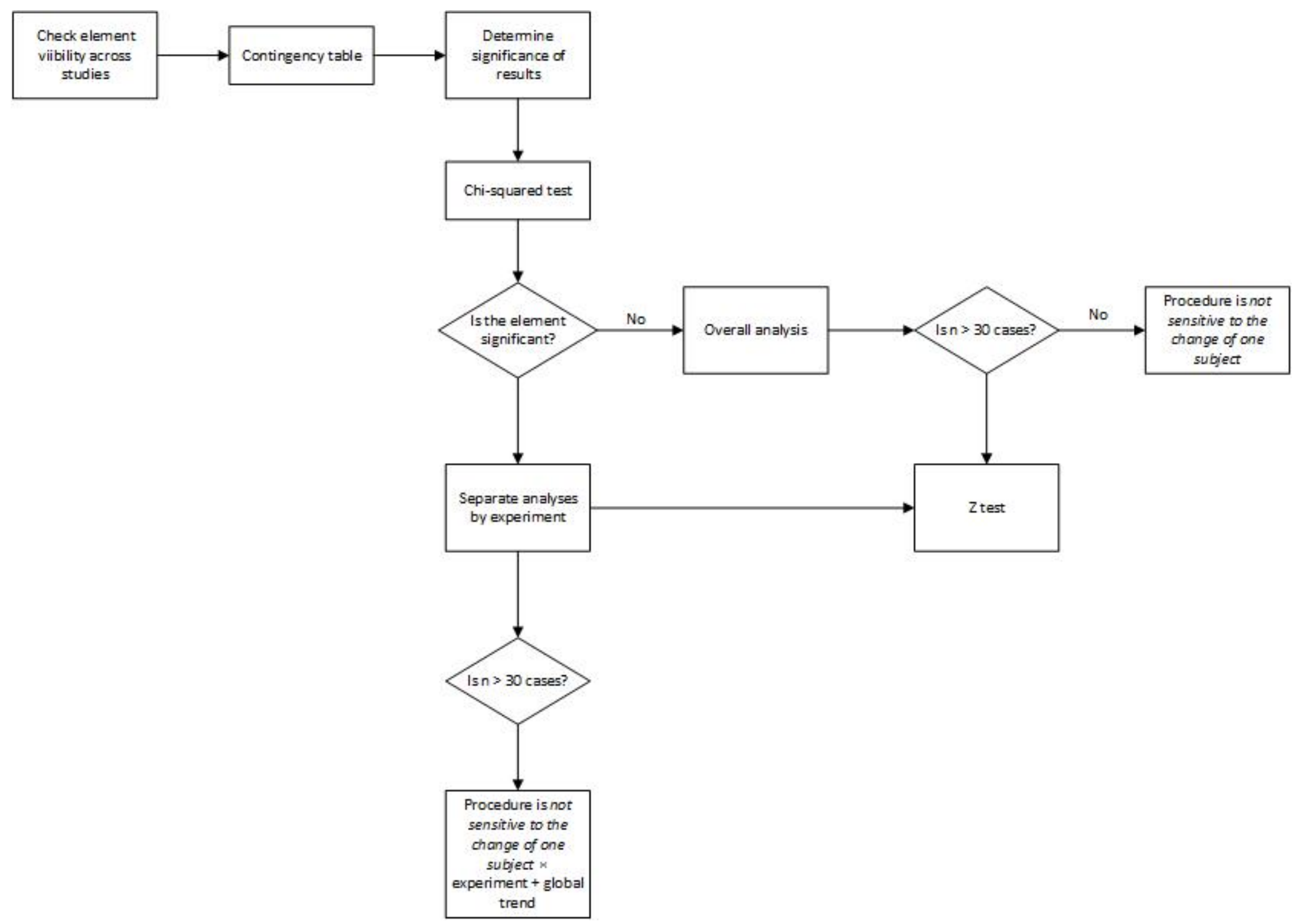

**Fig 1.** Steps taken to determine element visibility.

**Note that, instead of being asked to conduct an interview, the experimental subjects of the last two empirical studies (Q16a and Q16b) were given an audio describing the domain.** Please refer to Section 0 for details, but from the methodology viewpoint, we wish to clarify that, although the subjects did not get the chance to interact with the audio, that is, ask questions about the problem, they were at liberty to stop, rewind, or listen to parts of the audio again as often as they liked, provided that they did not go over the 20-minute time limit. At the end of the elicitation, they were given 25 minutes to answer a questionnaire on the problem domain using the information they could recall. We adopted this approach because our previous research revealed that, on average, students tended to identify about 50% of the problem domain elements. However, we had no way of knowing whether the identified information depended on the questions the analysts asked or the information the respondents provided. Using this approach, we ensure that all the domain-related information is mentioned.

### D. Operationalization of visibility

In this research, we use the concept of visibility. We operationalised the visibility of the domain elements as the **percentage of sessions (or analysts) that identify a specific problem domain element** (AP, IP). Using the above example, if 50% of the analysts identify the client concept during the analysis, client visibility would be 50%.

We used different methods and statistical tests to calculate and analyse visibility, including contingency tables,

probabilities and binomial distributions. Additionally, we built different charts and tables to ease the interpretation of the results. The steps taken to determine the level of element visibility shown in Fig 1 are detailed below.

We first determined whether the visibility of the elements depends on the empirical study in which it was measured. As we ran ten empirical studies, there was a possibility of some elements being visible in some experiments and not visible in others. To do this, we: a) used contingency tables to observe whether the number of times that a specified element is identified is consistent across studies, and b) applied the $\chi 2$ statistical test to determine the dependency ratio between element visibility and the experiment.

Using the contingency table, we can study the relationship between two or more variables and determine whether the marginal frequencies are similar. As far as we are concerned, this means that, for each element, this element is observed at a similar rate across studies or, alternatively, the visibility of the elements is unchanged across studies. The $\chi 2$ test tests the hypothesis that the variables under study are statistically independent, that is, element visibility does not depend on the study in which it was observed. If the result of the $\chi 2$ test is significant, there is a dependency between the study and element visibility, that is, the element is more visible in some studies than in others, whereas if the result is not significant, there is no dependency between element visibility and the study, that is, visibility does not depend on the study in which the element was elicited.

Knowing whether an element is significant within the contingency table is not enough to determine element visibility, and other procedures need to be applied to define the definite level of element visibility. Therefore, we applied a global between-studies analysis or a separate analysis for each experiment depending on the significance and population sample size, as shown in Fig 1. The global analysis consists of analysing the element visibility of all the studies as a whole, summarizing the experimental data of the family as if it were a single study. This is possible, as the studies are close replications of each other, where the subjects perform the same experimental task on the same problem domains. The separate analysis by study or within-study analysis determines the visibility of each element by study.

We operationalized the visibility level of an element using a 50% threshold. If an element is observed (or not observed) more (or less) than by chance, that is, 50%, in a set of interviews, it is considered to be visible (or not visible). We used the **Z-test** to determine the visibility level. Using the Z-test, we can make inferences on the probability of events occurring. In this case, the null and alternative hypotheses to be tested are the probability (P) of observing a specified element: $H_0$: = P = 0.5; $H_1$: P <> 0.5. This test cannot be applied unless the sample size is greater than 30 cases [31]. Note, however, that if this condition is met, this test can be applied in both the between- and within-studies analysis.

Due to the above sample size limitation, an alternative analysis procedure to the Z-test needs to be formulated when the sample size is smaller than 30. We refer to this alternative procedure as ***not sensitive to the change of subject.*** This criterion consists of comparing the number of subjects that do and do not observe the element to conclude when an element is visible, not visible, and undeterminable.

TABLE V summarizes when an element is visible, not visible, and undeterminable, depending on the size of the experimental population and the statistical significance of each of the methods or procedures specified in Fig 1. Appendix B provides examples illustrating how we applied each of the procedures.

## V. RESULTS

In response to the research question, we analysed the visibility of the elements making up the problem domains under analysis. We first analysed the IP domain, as it was the domain that generated the most empirical data. We then analysed the AP domain to check and confirm whether element visibility is consistent across different domains. The analysis procedure is described in detail in Appendix B. Note that, to calculate visibility, we used eight primary studies (from Q17 to E15) from the family of empirical studies shown in TABLE II, which, taken together, output the greatest number of experimental data. The last two studies (Q16a and Q16b) are analysed separately and used in the discussion.

The results are shown in the respective activity diagrams and concept models. The activity diagram represents the sequence of activities that comprise the typical domain process, whereas the concept model shows the key domain concepts and their relationship.

### *A. IP domain visibility*

The key characteristic of the IP domain is that it belongs to a domain with which all the experimental subjects acting as analysts were unfamiliar. Therefore, the identification of any element (and, therefore, its visibility level) should depend exclusively on interview performance or the execution of the experimental task, that is, it would depend on what actually happened during the elicitation session rather than on previous analyst experience.

The IP domain was explored in eight empirical studies by a total of 88 analysts who interviewed three different respondents. The results derived from the analysis should, therefore, be fairly generalizable, and, at the same time, some elements could be visible in some studies but not in others. In other words, element visibility could depend not on the problem but on the particular empirical study in which it was analysed. To check this, we applied the between-studies and within-studies analyses outlined in Section 4.4 and detailed in Appendices B.1.a and B.1.b, respectively.

The analysis result is shown in Fig. 2, illustrating the percentage visibility of the processes or activities (Fig. 2.a) and concepts (Fig. 2.b) identified by the experimental subjects. The visibility of the domain elements is reported as percentages and is colour coded: 1) the colour blue represents elements that are hard to observe, 2) the colour yellow denotes elements that are observed by chance, and 3) the colour orange stands for elements that are easily observable.

Fig. 2 shows that there are clusters composed of similarly visible elements. In the activity diagram (Fig. 2.a), activities (A2-A7) are visible and are clearly distinguishable from the others that are not visible. Note that 50% of the processes tend to be visible. The pattern in the concept model (Fig. 2.b) is similar. The visible concepts (C19-C22, C24) and (C1, C8-C12) are clearly distinguishable from undeterminable concepts (C13-C15), as well as the other concepts that are not visible. Note that 45.8% of the concepts are visible, 12.5% are undeterminable, and 45.8% are not visible.

The results suggest that **analysts tend to focus on particular problem-domain components (visible elements) and ignore others (non-visible elements)**.

TABLE V. PROCEDURES FOR CALCULATING THE VISIBILITY OF ELEMENTS

| Experimental Population Size | Elements | Procedure | Element Visibility | Condition | Observations |
| --- | --- | --- | --- | --- | --- |
| > 30 cases | **Significant and not significant** in contingency table | Z-test (Z) | Visible (V) | If Z is significant and P(v) > 0.5 | An element is visible if Z is significant and the probability of it being visible is greater than 0.5 (50%) |
| | | | Not visible (NV) | If Z is significant and P(nv) > 0.5 | An element is not visible if Z is significant and the probability of it not being visible is greater than 0.5 (50%) |
| | | | Undeterminable (Und) | If Z is not significant | An element is undeterminable when Z is not significant |
| < 30 cases | **Not significant** in contingency table | "not sensitive to change of subject" | Visible (V) | V > NV | An element is visible if the number of visible elements is greater than the number of non-visible elements. |
| | | | Not visible (NV) | NV > V | An element is not visible if the number of non-visible elements is greater than the number of visible elements. |
| | | | Undeterminable (Und) | $\lvert V – NV \rvert$ <= 2 | An element is undeterminable when switching a subject from the N to the NV set (or vice versa) does not change majority. |
| | **Significant** in contingency table | "not sensitive to subject change"+ global trend | Visible (V) | Visibility is determined: i) by each individual experiment; ii) the global visibility according to the observed trend across the experiments. | An element is visible if element visibility is visible in at least 4 out of the 6 experiments. |
| | | | Not visible (NV) | | An element is not visible if element visibility is not visible in at least 4 experiments. |
| | | | Undeterminable (Und) | | It is not possible to determine whether or not the element is visible. Visibility is heterogeneous across the series of experiments. |

### *B. AP domain visibility*

Unlike the IP domain, subjects are familiar with the AP domain. We analysed four replications, totalling 36 experimental subjects and two respondents. The analysis procedure is detailed in Appendix B.2. Fig. 3 shows the percentage visibility of the processes or activities (Fig. 3.a) and concepts (Fig. 3.b) identified by the experimental subjects. The same colours are used in Fig. 3 to mean the same as in Fig. 2. In the activity diagram shown in Fig. 3.a, activities (A1, A3, A4, A10, and A16) are undeterminable, whereas all the other elements are non-visible, except for A9, which is visible, that is, 31.25% of the processes tend to be undeterminable, 62.5%, non-visible and 6.25%, visible. For the concept model (Fig. 3.b), we found that concepts (C1, C2, C3, C4 and C5) are visible, whereas C7 and C9 are not visible and C6, C8 and C10 are undeterminable, that is, 50% tend to be visible, 20%, non-visible and 30%, undeterminable. Again, **subjects tend to focus on specific problem-domain components.**

## VI. OTHER ISSUES THAT COULD INFLUENCE ELEMENT VISIBILITY

Kiris [17] suggests that the personal characteristics of the experts and knowledge engineers influence elicitation effectiveness. The requirements community probably agrees that such characteristics can influence the course of a conversation and, therefore, the visibility of the domain elements. In our research, we were able to study analyst experience and its impact on respondents.

In this research, we defined the visibility of a specified element according to the behaviour of a set of analysts that participated in several experimental replications. As shown in TABLE II, the characteristics of the analysts that participated in the replications are wide ranging (students and practitioners with varying years of experience). As a result, element visibility could depend on *analyst experience*, that is, some elements might not have been visible because inexperienced analysts (most of the subjects) tend to overlook elements, whereas some, if not all, of these elements that inexperienced analysts find hard to pick out, would be visible to experienced analysts. As described in Appendix C, our results highlight the fact that **experience does not appear to affect element visibility**. Subjects tend to identify problem domain elements similarly, that is, element visibility is unchanged as analyst experience increases. On the other hand, we confirm that changes in element visibility (both from a global perspective and on an experience-based scale) are very limited.

*A. Analyst experience-dependent element visibility*

To analyse respondent influence, we applied the same visibility analysis procedures as used previously to each respondent separately, as specified in Appendix D. The results show that the visibility of the domain elements largely depend on the respondent, but visibility is mediated by the domain type:

- In the IP domain, there are hardly any discrepancies between the respondents (OD, JW, AG). Whenever there are discrepancies, they tend to be gradual, that is, an element that is hard to observe for some respondents is easily observed by others and by chance by others.
- In the AP domain, there are notable discrepancies among respondents. Generally, element visibility is greater when the respondent is JW and lower when the respondent is OD.

One reason that might explain the difference between respondents is the language. As shown in TABLE II, OD and AG use English (a foreign language for both) and JW uses Spanish (his mother tongue). However, language does not appear to be a completely convincing explanation. All three respondents behave similarly for IP, whereas the respondent JW is more communicative than OD for AP. Thus, the key would appear to be the domain rather than the language. We should recall that analysts are unfamiliar with the IP domain, whereas they are familiar with AP domain. An explanation could be that analysts need to invest more time and effort in understanding the unfamiliar domain. As this takes more time and effort, respondent verbosity has little influence on the results, as the respondents tend to speak less and always on the same topics. When the domain is familiar, either respondent verbosity or language fluency influences element visibility.

**A. Activities**

**B. Concepts**

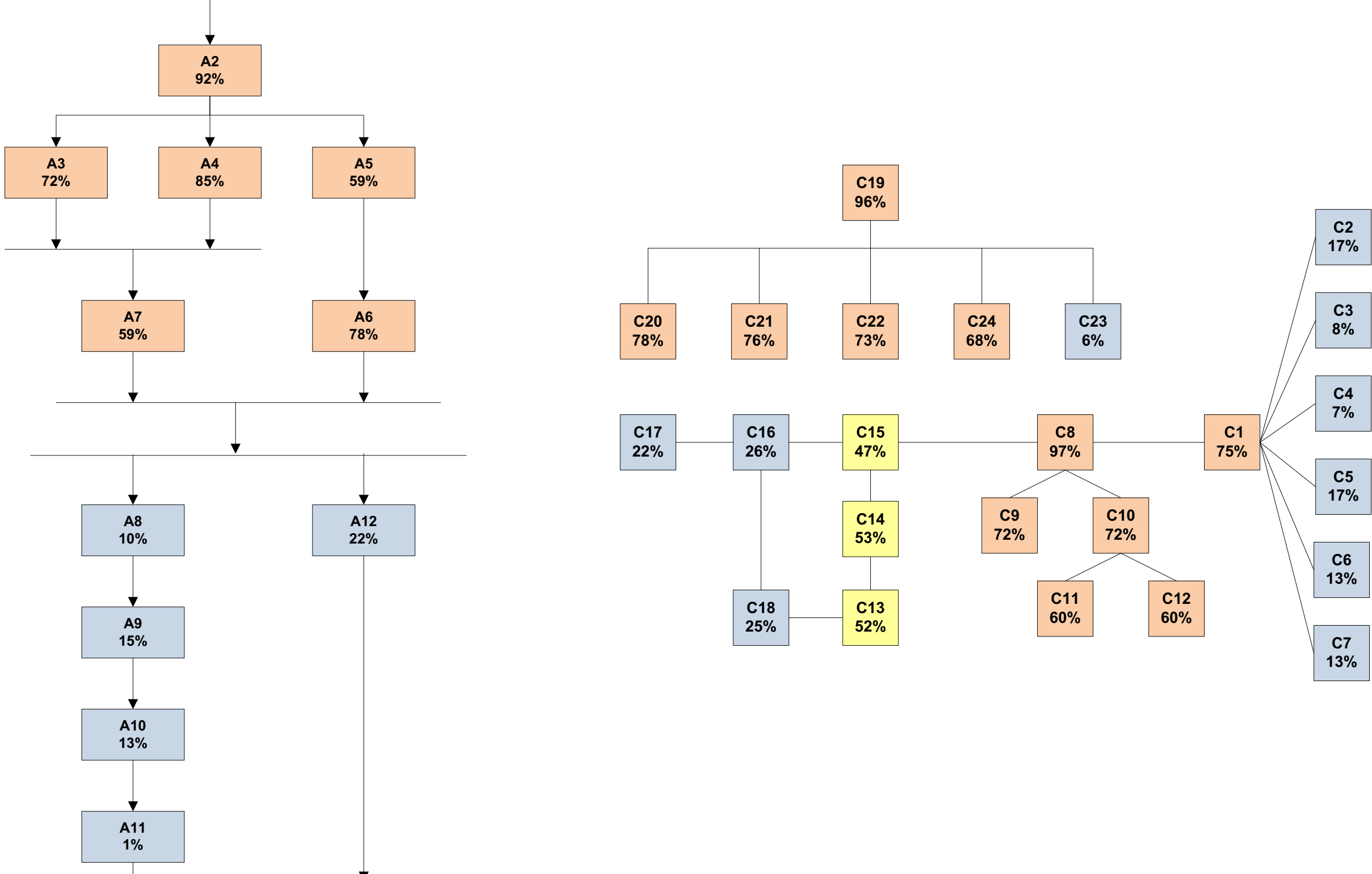

**Fig. 2.** IP element visibility: (a) activity diagram; (b) concept model

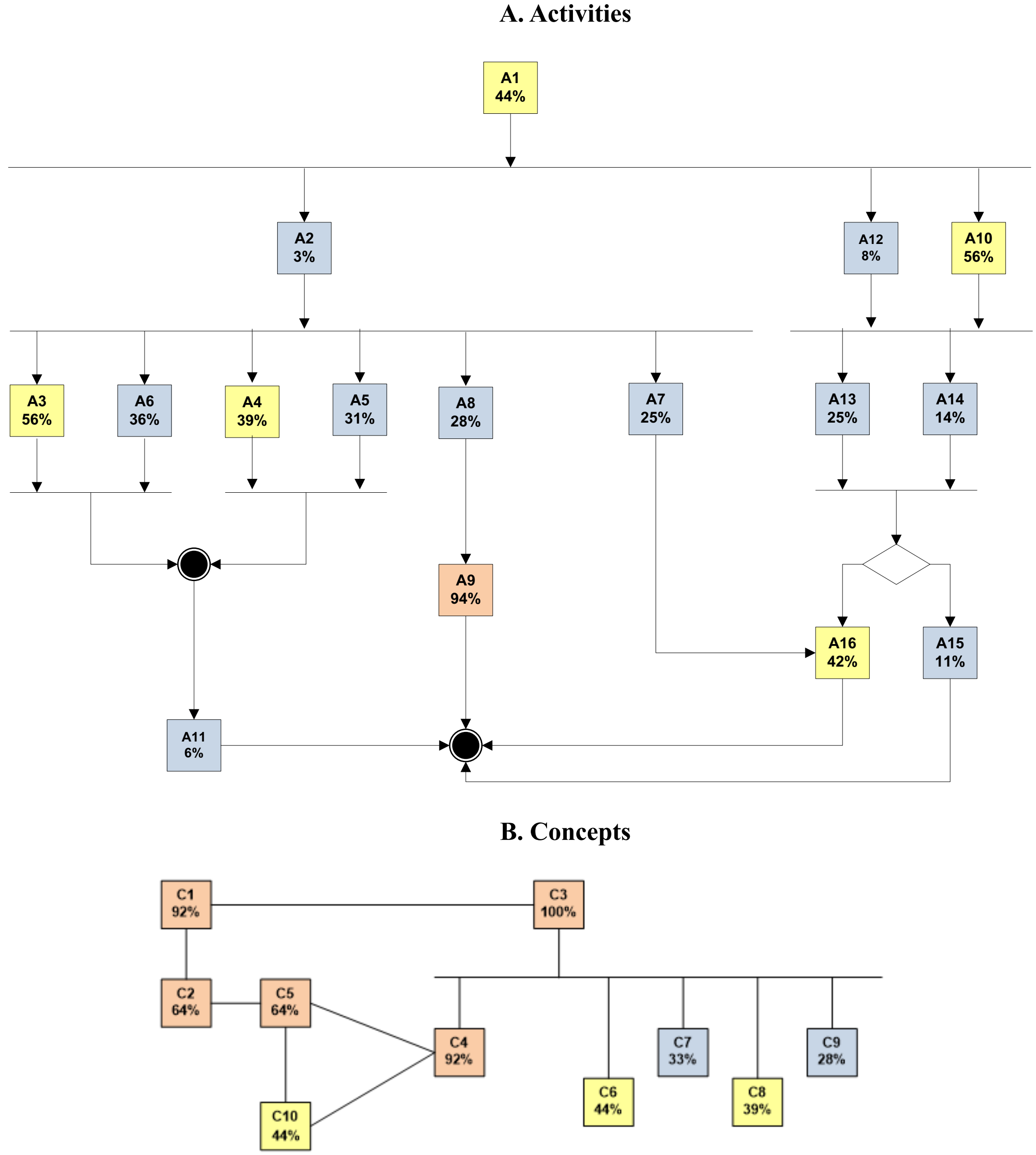

**Fig. 3.** AP element visibility: (a) activity diagram; (b) concept model

## VII. Discussion

Our results show that, irrespective of the analysed problem domain, subjects tend to focus on a subset of the problem domain elements during the requirements elicitation process. In the following, we try to explain the reasons for these differences drawing on the related work.

### *A. Domain influence*

Before discussing the more speculative aspects, we should note at this point that there are differences between the IP and AP domains with respect to the number of visible elements. TABLE VI summarizes these differences. In the IP domain, the activities are more visible than in the AP domain, but the opposite applies to the concepts. If we consider the sum of the concepts and activities, the elements appear to be more visible in the IP than in the AP domain [17][25], that is, analyst effectiveness varies depending on the domain.

However, these data should be viewed with caution, as many more interviews were conducted with IP than with AP, and, therefore, the IP data are more reliable than the AP data. Note that the number of undeterminable elements in AP is much greater than in IP. If we were to add up the undeterminable and visible elements in each domain, the percentages of visibility (and non-visibility) for IP and AP would be very similar.

TABLE VI. IP AND AP DOMAIN ELEMENT VISIBILITY

| | | IP | AP |
|---|---|---|---|
| **Activities** | Visible | 50% | 6% |
| | Undeterminable | 0% | 31% |
| | Non-visible | 50% | 63% |
| **Concepts** | Visible | 46% | 50% |
| | Undeterminable | 13% | 30% |
| | Non-visible | 42% | 20% |
| **Total** | Visible | 47% | 23% |
| | Undeterminable | 8% | 31% |
| | Non-visible | 44% | 46% |

### *B. Influence of domain complexity*

Another striking aspect is the fact that, a priori, the visibility results contradict what we thought that we knew about the influence of domain knowledge: analysts appear to be more effective in familiar domains [15]. In our research, IP is an unfamiliar domain, which means that the visibility of the AP elements should be greater, which is precisely the opposite of what we observed.

However, these findings can be reconciled. As shown in Fig. 2.a and Fig. 3.a, the IP activity diagram is much simpler than the AP diagram (the complexity of the concept diagrams is comparable). We hypothesize that it is this complexity that could be behind the smaller number of visible elements in AP. There is another reason that justifies this hypothesis. Concluding (like other works [32]) that analyst effectiveness increases in familiar domains, Aranda et al. [15] uses an experimental design that counterbalances domain complexity by using a similar activity diagram as shown in Fig. 2.a for a familiar domain and another similar to the one shown in Fig. 3.a for an unfamiliar domain. There is no such counterbalancing in this research (the domain is always simple for IP and complex for AP), which could explain the lower visibility of the AP elements. In any case, we reaffirm that, for the time being, this is a mere hypothesis.

### *C. Influence of instrumental factors*

The personal characteristics of the analyst, such as domain knowledge or experience, could be other factors explaining element visibility. However, our findings suggest that whether or not **the analyst identifies a particular domain element does not appear to depend on such factors.**

The respondent's articulateness (i.e., whether he or she provides more or less information) does appear to affect element visibility, although the impact is only clearly perceived in the AP domain. In IP domain, visibility depends more on the analyst than on the respondent.

Visibility could be related to the number and type of questions asked by the analyst and his or her ability to take notes about the interview. Therefore, **we asked ourselves what would happen if we eliminated the analyst-client conversation, that is, if the analyst were to be given all the domain elements using an audio, and if, instead of using a written report, analysts were asked to complete a questionnaire asking about the domain elements**. In both cases, the problem would be described in full, but there would be no interaction between the analyst and respondent to limit the number of domain elements that the analyst is able to identify. Besides, we would assure, in the second case, that the report prepared by the analyst did not omit any element that he or she had identified from the audio due to forgetfulness or time pressure, etc.

Studies Q16a [33] and Q16b [34] investigate precisely these conditions. Q16a studies the IP domain, whereas Q16b examines the AP domain. In both cases, analysts are given an audio describing the problem domain as a whole and then asked first to produce a written report and then to fill in a questionnaire (to prevent the questionnaire from "jogging" the memory of the analyst). Otherwise, Q16a and Q16b are internal replications of E12 that use the same description of the AP and IP domains specified in Appendix A.

Fig. 4 shows a box plot in which the y-axis represents the mean effectiveness of experimental subjects, and the x-axis represents the elicitation technique (interview or audio) and the document (report or questionnaire) used to measure subject effectiveness. Effectiveness is calculated as in [9] and refers to the total percentage of elements identified by subjects.

Our results (detailed in Appendix E) suggest that there is a slight improvement in subject effectiveness when using audio as an elicitation technique and a more marked improvement when using the questionnaire after requirements elicitation, as shown in Fig. 4. Effectiveness is linked only indirectly to domain element visibility, which suggests that the audio plus written report will not improve visibility very much, but **subject effectiveness improves notably with the application of the questionnaire.**

After listening to the audio, subjects identified 35.19% and 71% of the domain elements in the reports and questionnaires, respectively, for the AP domain, whereas the rate was 56% and 67% for the IP domain. Note that a comparison of reporting by subjects participating in the experiment series (Q07 to E15) with Q16 subjects shows that average effectiveness is unchanged for AP (35%) and drops slightly for IP (42%). Requirements elicitation is limited to the information provided by the audio and the information that the analyst was able to record throughout the process. However, the questionnaire helps analysts to recall the captured information. This is reflected in our results, as elements (processes and concepts) tend to be more visible. Fig. 6 illustrates the example of IP element visibility, grouped by elicitation technique (interview or audio) and information acquisition method (report or questionnaire). All the analyses are available in Appendix E.2.a, whereas Appendix E.2.b contains the analyses for AP. Fig. 6 clearly shows that the use of audio increases element visibility. In fact, elements that tended to be undeterminable or not visible in the experiment series (Q07 to E15) either become visible or are more visible in Q16: C15, C14, C13, C2-C3, etc. The difference is even bigger using the questionnaire (Q16Audio-Questionnaire).

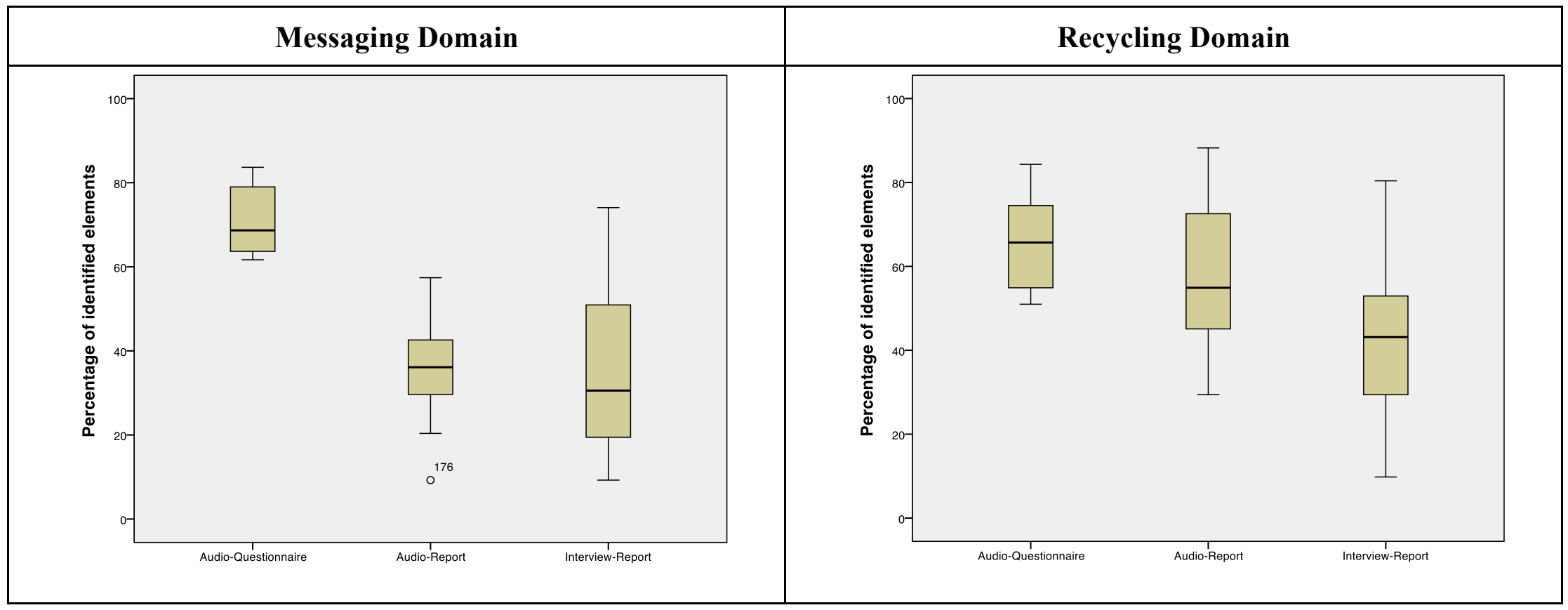

**Fig. 4.** Subject effectiveness grouped by problem domain and elicitation technique

In our opinion, these findings suggest that the cause of low element visibility is not the poor quality of the analyst-client conversation. Note that, were the interview to be the cause of low visibility, we should observe a notable increase in the effectiveness of the audio+report experimental condition, which is not the case. Therefore, it appears to be the report that is at fault, where analysts fail to write down all the knowledge that they have acquired [35]. An incomplete report can lead to ambiguities and/or omissions in the elaboration of the software requirements specification.

## D. Influence of focal areas

To conclude the discussion, Fig. 5 offers an alternative perspective of the results, which is probably related to the areas on which the analysts focus during the interviews.

Fig. 5 (see also Figs. 15-18 in Appendix G) show exactly the same information as Fig. 2 and Fig. 3. However, they use iso-probability curves to highlight the domain elements that have the same probability (in intervals of 10%) of being observed. The colours used are also related to visibility. The darkest blues represent the least visible elements, whereas the darkest oranges represent the most visible elements. Figures 6 to 9 are essentially colour-coded maps to highlight the most or least visible elements.

The most striking feature is that the visible elements form clusters. There is substantial evidence from programming and design to the effect that mental schemas exist [26]. It should, therefore, come as no surprise that they exist in software requirements field too. Additionally, we also observe that these clusters are formed based on a central element. In all cases, this is a vital element in the respective domain. More specifically, the most visible elements (higher levels) of the IP domain are:

- C8: Battery
- C19: Machine
- A2: [A machine] Separate batteries

whereas for the AP domain, the most visible elements are:

- C1: User
- C3: Messages
- A9: [A user] Send messages
- A10: [A user] Receive messages.

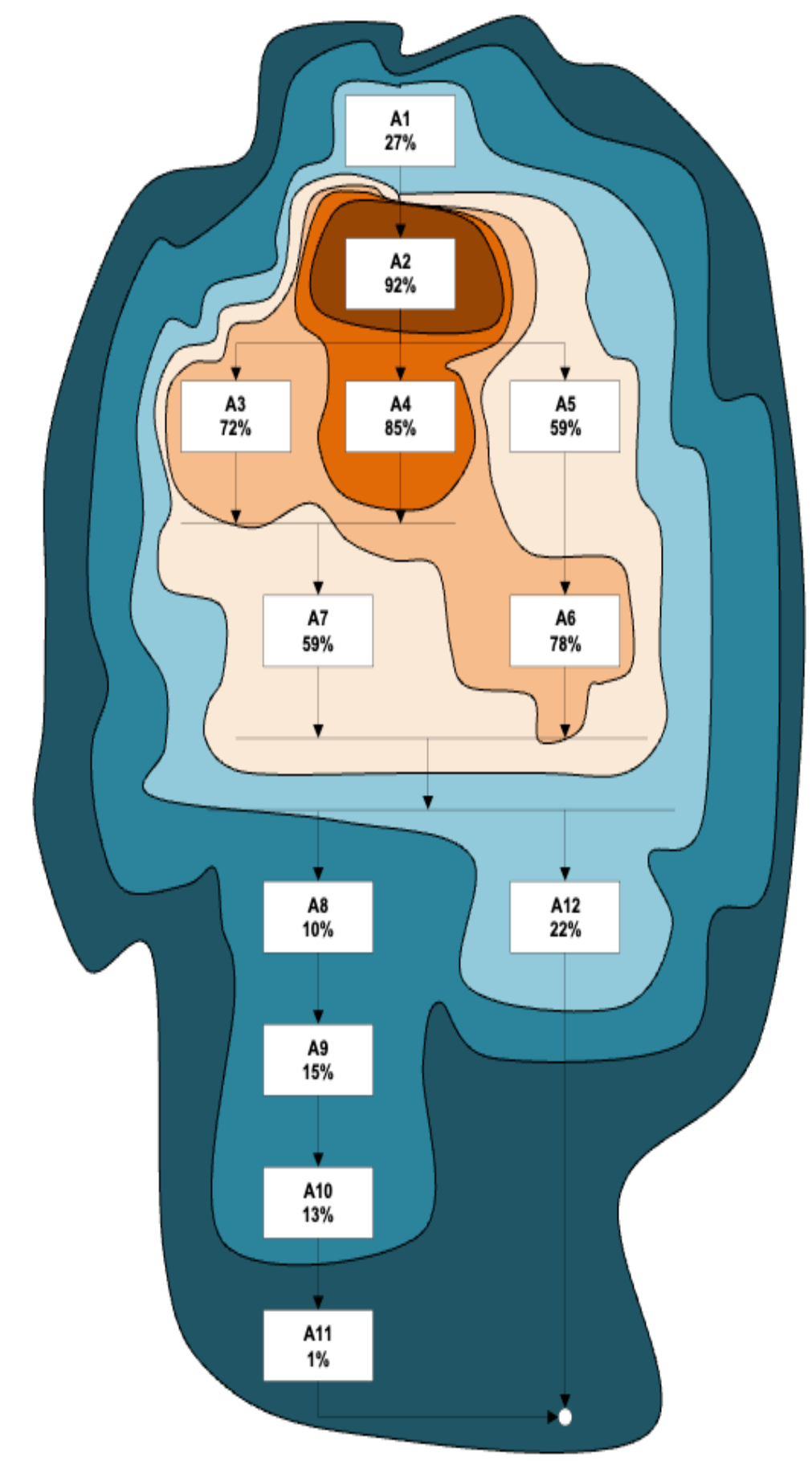

**Fig. 5.** Iso-probability curves for IP: Activity diagram

The importance of these elements is self-evident. Based on these elements, Fig. 5 (and the figures in the appendix) clearly illustrate that there is a gradient that, in our opinion, is related to how the concepts are acquired by the analysts. In other words, the analysts appear to start the interview by identifying the most relevant domain elements and then join the dots by describing related elements. The literature on psychology and education, like [36], supports this approach. However, we will not go any further down that road because this is not the place to provide the theoretical groundwork for what is a mere working hypothesis.

One question that requires further attention is why analysts stop at the clusters shown in Fig. 5. To be honest, we believe that analyst progress is not as bad as it looks: analysts identify, on average, almost 50% of the domain elements (see TABLE VI). This is quite good, taking into account that they have performed only one short interview. Having said this, we hypothesize that the non-visible areas of Fig. 5 contains elements that are far removed, in terms of relations and prelations, from the visible elements. Analysts are somehow unable to skip from the clusters of visible elements to non-visible elements. This is, on the other hand, logical from the viewpoint of some learning theories, like meaningful learning [36], for example.

In our view, the related work, reported in Section II, provides additional support for this hypothesis. Analysts who were familiar with the domain, because either they have experience in or the interview types focus on specific domains, could broaden the focus of their inquiry. This interpretation would be perfectly consistent with the results reported by Agarwal and Tanniru [16], and Kiris [17]. The use of different types of prompts would have a similar effect, that is, they would enable the analysts to change the focus of the conversation, which should improve effectiveness. This interpretation is matches the findings of Wilt [18] Browne and Rogich [22] and Pitts and Browne [23].

## VIII. VALIDITY THREATS

### A. Statistical conclusion validity threats

**Measurement reliability.** Visibility is operationalized depending on the number of subjects that have or have not identified the domain element, that is, an element may or may not be *visible* (if there are probabilities greater or less than 50% of it being identified). Note, however, that an element may be *undeterminable* (that is, when 50% of the subjects have or have not identified the element or when there are one or more subjects that can alter the visible or non-visible trends). This largely depends on the number of subjects or sample size. Separate experiment analysis, experience categorization or respondent grouping all have a bearing on sample size.

**Sample size.** To analyse the influence of audio and questionnaires on the effectiveness of analyst elicitation, and therefore on the visibility of the elements defining the analysed domains, we used two quasi-experiments run in 2016 (Q16). These experiments are not part of the overall analysis of visibilities (as they use different methodologies). These experiments had 14 subjects for the IP domain and ten for the AP domain, which could influence the respective analyses.

### B. Internal validity threats

**Requirements elicitation using audio.** The number of times the analyst listened to the audio describing the problem domain. A time limit of 20 minutes was placed on the activity of listening to the audio containing the description of the domain for an audio with a duration of 6 minutes and 42 seconds.

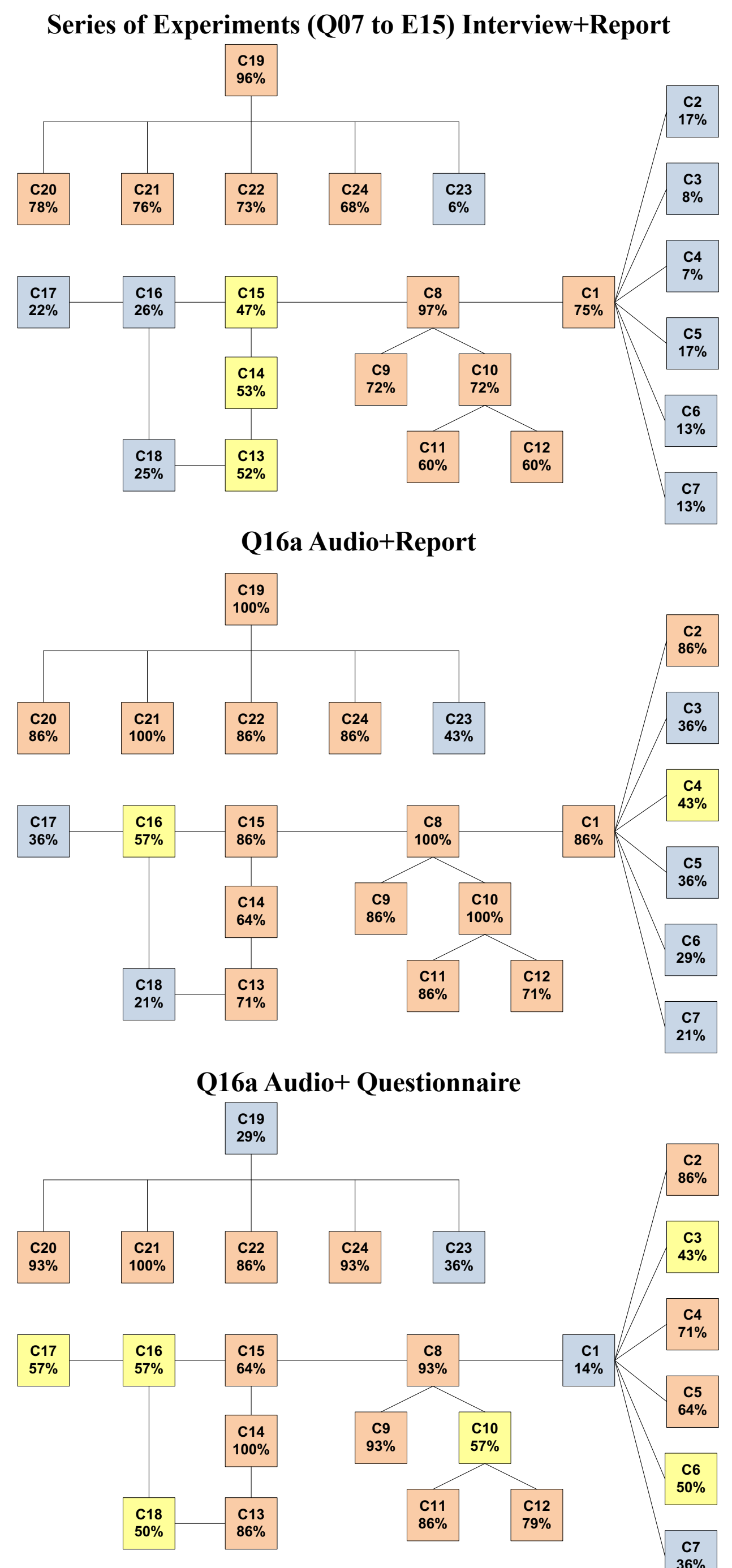

**Fig. 6.** Concept visibility in IP domain, grouped by information acquisition method.

### *C. Construct validity threats*

**Dichotomization or experience categorization.** To study the effect of experience on element visibility, we grouped experience into three subgroups: novice (0-1 years), intermediate (2-4 years), and expert (>=5 years). When working with subgroups, the sample size decreases, thus lowering the accuracy of the domain element visibility estimation (probability of observation). While dichotomization or categorization is potentially a useful method for group comparisons and improves the interpretability of the results, the results are less reliable due to information loss with respect to the raw data.

**Visibility operationalization:** We applied a rather complex analysis procedure because of the particularities of the analysed family of experiments. A simpler procedure, depending exclusively on the Z-test, for example, might have output slightly different results.

### *D. External validity threats*

**Generalization:** The fact that the experimental subjects come from a convenience rather than a randomized sample (that is, subjects are students enrolled in a specified course that have not been recruited from a larger population) poses a threat to the external validity of the experiment. To deal with this threat, it would be necessary to replicate the studies on more heterogeneous populations.

## IX. Conclusions

In order to gain a better understanding of the process of how analysts capture information, we conducted an exploratory study of visibility (capture or identification by analysts) of the elements defining the problem domain. We categorized the visibility of the elements: *easy to observe* (usually identified by subjects), *hard to observe* (normally overlooked by the subjects) and *observed by chance* (sometimes identified and sometimes overlooked).

To do this, we used a well-known problem domain (AP) and an unfamiliar problem domain (IP) for elicitation by requirements analysts. We conducted ten empirical studies with students from the Universidad Politécnica de Madrid as part of a requirements engineering course, and with professionals recruited at the REFSQ conference.

Our results highlight that subjects tend to focus on a subset of elements, irrespective of the problem domain. Within this data subset: a) subjects tend to focus more on the processes in unfamiliar domains, and b) more on concepts in familiar domains. On the other hand, we tried to empirically check factors that could influence the visibility of these elements, such as analyst experience, respondents, as well as the information acquisition method (audio, questionnaire). The observed trends suggest that:

- Experience (novice, subjects with intermediate experience and experts) **does not appear to have an impact on visibility.**
- Respondent (OD, JW, AG) tends to lead to very similar element visibility (with some exceptions) in the unfamiliar domain, where **we observed slightly more visibility than in the familiar domain**.
- Audio improves subject effectiveness, and, therefore, more elements tend to be visible.
- Questionnaire improves subject effectiveness and element visibility considerably (in both familiar and unfamiliar domains).
- Report **only slightly improves effectiveness and visibility** when analysts listen to the domain description in both the familiar and unfamiliar domain.

It is possible, although this is only a hypothesis at this stage, that domain element visibility is also influenced by:

- **Domain complexity**, particularly by relationships between different elements.
- **Relative importance** of the different elements within the domain, and
- **Interview script**, possibly because the questions include some domain knowledge and/or it uses prompts to elicit knowledge from respondents.

Obviously, this exploratory study has only scratched the surface of the problem of how analysts capture the domain elements. Further research is required to replicate our results. However, we believe that it is interesting to have observed that the visibility problem may be caused not by the analyst-client conversation but by the subsequent report. This signposts a potentially promising line of research to explain how to acquire and document requirements elicitation information.

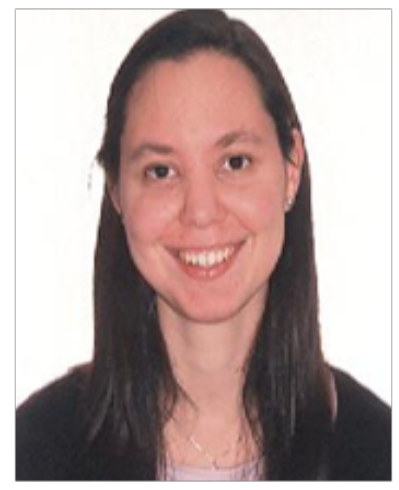

**Alejandrina M. Aranda** received the MSc degree in Information Technology and the PhD degree from the Universidad Politécnica de Madrid. She is a researcher with the UPM's School of Computing. Her research interests include Requirements Engineering and Empirical Software Engineering.

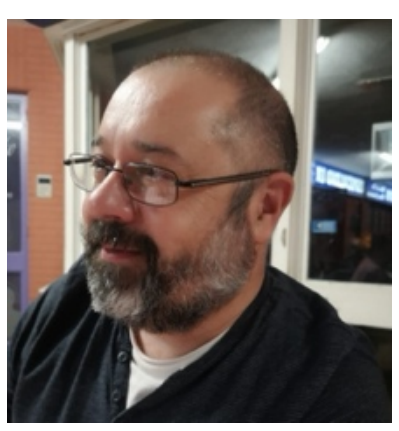

**Oscar Dieste** received his BS and MS in Computing from the University of La Coruña and his PhD from the University of Castilla-La Mancha. He is a researcher with the UPM's School of Computer Engineering. He was previously with the University of Colorado at Colorado Springs (as a Fulbright scholar), the Complutense University of Madrid, and the Alfonso X el Sabio University. His research interests include Empirical Software Engineering and Requirements Engineering.

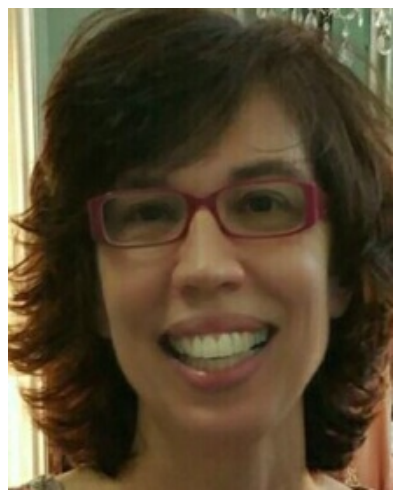

**Natalia Juristo** is *full professor* of software engineering with the Computing School at the Technical University of Madrid (UPM) since 1997 and holds a FiDiPro (Finland Distinguish Professor) research grant since 2013. She was the Director of the UPM MSc in Software Engineering from 1992 to 2002 and the coordinator of the Erasmus Mundus European Master on SE (whith the participation of the University of Bolzano, the University of Kaiserslautern and the University of Blekinge) from 2007 to 2012.Natalia has served in several *Program Committees* ICSE, RE, REFSQ, ESEM, ISESE and others. She has been *Program Chair* EASE13, ISESE04 and SEKE97 and General Chair for ESEM07, SNPD02 and SEKE01. She has been member of several Editorial Boards, including Transactions on SE, Journal of Empirial Software Engineering and Software magazine. Dr. Juristo has been Guest Editor of special issues in several journals, including Journal of Empirical Software Engineering, IEEE Software, Journal of Software and Systems, Data and Knowledge Engineering and the International Journal of Software Engineering and Knowledge Engineering.